\documentclass[twocolumn]{aastex631} 
\usepackage{amsthm,amssymb,mathtools}
\usepackage{natbib}
\usepackage{multirow}
\usepackage{physics}
\usepackage{gensymb}
\usepackage{xcolor}
\usepackage{graphicx} 
\usepackage{subfigure}

\begin{document}
	\accepted{December 2, 2021}
	\published{February 28, 2022}
	\submitjournal{ApJS}
	\title{The OGLE Collection of Variable Stars. One Thousand Heartbeat Stars in the Galactic Bulge and Magellanic Clouds}
	\author[0000-0002-3051-274X]{Marcin Wrona}
	\affil{Astronomical Observatory, University of Warsaw, Al. Ujazdowskie 4, 00-478 Warszawa, Poland}
	\author[0000-0002-3218-2684]{Milena Ratajczak}
	\affil{Astronomical Observatory, University of Warsaw, Al. Ujazdowskie 4, 00-478 Warszawa, Poland}
	\author[0000-0003-2244-1512]{Piotr A. Ko\l{}aczek-Szyma\'nski}
	\affil{Astronomical Institute, University of Wroc\l{}aw, Kopernika 11, 51-622 Wroc\l{}aw, Poland}
	\author[0000-0003-4084-880X]{Szymon Koz{\l}owski}
	\affil{Astronomical Observatory, University of Warsaw, Al. Ujazdowskie 4, 00-478 Warszawa, Poland}
	\author[0000-0002-7777-0842]{Igor Soszy{\'n}ski}
	\affil{Astronomical Observatory, University of Warsaw, Al. Ujazdowskie 4, 00-478 Warszawa, Poland}
	\author[0000-0002-6212-7221]{Patryk Iwanek}
	\affil{Astronomical Observatory, University of Warsaw, Al. Ujazdowskie 4, 00-478 Warszawa, Poland}
	\author[0000-0001-5207-5619]{Andrzej Udalski}
	\affil{Astronomical Observatory, University of Warsaw, Al. Ujazdowskie 4, 00-478 Warszawa, Poland}
	\author[0000-0002-0548-8995]{Micha\l{} K. Szyma{\'n}ski}
	\affil{Astronomical Observatory, University of Warsaw, Al. Ujazdowskie 4, 00-478 Warszawa, Poland}
	\author[0000-0002-2339-5899]{Pawe\l{} Pietrukowicz}
	\affil{Astronomical Observatory, University of Warsaw, Al. Ujazdowskie 4, 00-478 Warszawa, Poland}
	\author[0000-0001-9439-604X]{Dorota M. Skowron}
	\affil{Astronomical Observatory, University of Warsaw, Al. Ujazdowskie 4, 00-478 Warszawa, Poland}
	\author[0000-0002-2335-1730]{Jan Skowron}
	\affil{Astronomical Observatory, University of Warsaw, Al. Ujazdowskie 4, 00-478 Warszawa, Poland}
	\author[0000-0001-7016-1692]{Przemek Mr\'oz}
	\affil{Astronomical Observatory, University of Warsaw, Al. Ujazdowskie 4, 00-478 Warszawa, Poland}
	\author[0000-0002-9245-6368]{Rados\l{}aw Poleski}
	\affil{Astronomical Observatory, University of Warsaw, Al. Ujazdowskie 4, 00-478 Warszawa, Poland}
	\author[0000-0002-1650-1518]{Mariusz Gromadzki}
	\affil{Astronomical Observatory, University of Warsaw, Al. Ujazdowskie 4, 00-478 Warszawa, Poland}
	\author[0000-0001-6364-408X]{Krzysztof Ulaczyk}
	\affil{Astronomical Observatory, University of Warsaw, Al. Ujazdowskie 4, 00-478 Warszawa, Poland}
	\affil{Department of Physics, University of Warwick, Gibbet Hill Road, Coventry, CV4 7AL, UK}
	\author[0000-0002-9326-9329]{Krzysztof Rybicki}
	\affil{Astronomical Observatory, University of Warsaw, Al. Ujazdowskie 4, 00-478 Warszawa, Poland}
	
	\correspondingauthor{Marcin Wrona}
	\email{mwrona@astrouw.edu.pl}

	\begin{abstract}
		We present a collection of 991 heartbeat star (HBS) candidates found in the Optical Gravitational Lensing Experiment (OGLE) project data archive. We discuss the selection process of the HBS candidates and the structure of the catalog itself. It consists of 512 stars located toward the Galactic bulge, 439 stars located in the Large Magellanic Cloud, and 40 in the Small Magellanic Cloud. The collection contains two large groups of HBSs with different physical properties. The main distinction between the two groups is the evolutionary status of the primary star. The first group of about 100 systems contains a hot main-sequence or a Hertzsprung-gap primary star, while the second group of about 900 systems includes a red giant. For each star, we provide two-decade-long time-series photometry, in the Cousins $I$- and Johnson $V$-band filters, obtained by the OGLE project. We also present basic observational information as well as orbital parameters derived from the light-curve modeling.
	\end{abstract}
	
	\section{Introduction}
	Heartbeat stars (hereafter HBSs) are a subclass of ellipsoidal binaries on eccentric orbits. In these systems, the brightness variations are caused by a tidal deformation of the components and by other proximity effects. The strongest changes of the total flux occur during the periastron passage. The name of this type of variable stars refers to a characteristic shape of the light curve, which is similar to an electrocardiogram signature. 
	
	The HBSs were introduced as a new class of variables by \cite{2012ApJ...753...86T}. They described the discovery of 17 HBSs in the Kepler space mission data archive (\citealt{2010ApJ...713L..79K}). For each system, the primary is the main-sequence (MS) star of spectral type from A to G, while the secondary is not visible or barely visible in spectral lines. 
	
	Before the publication of the catalog of HBSs by \cite{2012ApJ...753...86T}, only few such systems had been discovered and comprehensively described (e.g., among B-type stars; \citealt{2000A&A...355.1015D}; \citealt{2002A&A...384..441W}; \citealt{2009A&A...508.1375M}; and among A-type stars; \citealt{2002MNRAS.333..262H}). One of the most studied HBSs is KOI-54, which is considered as the archetype of this class (e.g., \citealt{2011ApJS..197....4W}; \citealt{2012MNRAS.420.3126F}). KOI-54 consists of two very similar A-type near-MS stars, on a highly eccentric orbit ($e=0.83$). The observed radial-velocity curve for this system proves that the most prominent brightening takes place when the stars are close to the periastron.
	
	The first large catalog of the Kepler HBSs was created by \cite{2016AJ....151...68K}. They presented, among other variable stars, a sample of 173 HBSs with short orbital periods (a median value of about 14 days) and very small amplitudes of brightness variations (mostly less than one millimagnitude). The classification of these systems as HBSs was mainly based on the shape of their light curves. Only a few of these objects have been studied in detail. The majority of the Kepler HBSs are low- and intermediate-mass MS stars of type A or F.
	
	Recently, \cite{2020arXiv201211559K} presented a collection of 20 new massive HBSs found in the Transiting Exoplanet Survey Satellite (TESS) data. Their sample of HBSs includes objects of spectral types from O to F (mainly A type), with orbital periods from a few days to about 25 days. 
	
	The phenomenon of heartbeat variability has been also found in systems with red-giant (RG) stars. In the Optical Gravitational Lensing Experiment (OGLE) catalog of ellipsoidal variables (\citealt{2004AcA....54..347S}), authors distinguished a sample of more than 100 stars with asymmetric light curves, contrary to sinusoidal-like shapes for "classical" ellipsoidal binaries. \cite{2004AcA....54..347S} proposed that such deviations are the result of tidal distortions and large eccentricity of the systems. A sample of 22 of these objects was examined in more detail by \cite{2017ApJ...835..209N}. They carried out simultaneous modeling of the light and radial-velocity curves using the Wilson–-Devinney code (\citealt{1971ApJ...166..605W}). Further, a group of 18 RG HBSs from the Kepler data were described by \cite{2014A&A...564A..36B}.
	
	In this work, we present a catalog of 991 candidates for HBSs, found in the OGLE database. The catalog consists of 512 systems located toward the Galactic bulge (GB), 439 systems located in the Large Magellanic Cloud (LMC), and 40 in the Small Magellanic Cloud (SMC). In Section 2, we describe the photometric data which were used in the analysis. In Section 3, we present the method of selecting candidates for HBSs. In Section 4, we discuss the results of the basic analysis of our HBS sample and show example light curves. In Section 5, we describe the structure of the catalog itself and how to access the data. In Section 6, we conclude our work.

	\section{Photometric Data}
	In this work, we use data collected mainly during the fourth phase of the OGLE project (OGLE-IV), which lasted from 2010 March until 2020 March, when the COVID-19 pandemic forced the OGLE operations to stop. All the observations were taken using the 1.3 m Warsaw telescope, which is located at Las Campanas Observatory, Chile (the facility site is operated by the Carnegie Institution for Science). During the OGLE-IV phase, the telescope was equipped with a mosaic camera consisting of 32 2K$\times$4K CCD sensors. The total field of view was about 1.4 square degrees and the pixel scale was $0.26''$.
	
	The vast majority of photometric data were collected in a Cousins $I$-band filter. The magnitude range achievable with the Warsaw telescope for this passband is from 13~mag up to 21.5~mag. Fields located in the GB were observed with an exposure time of 100 seconds. The number of observations strongly depends on the individual field: the least-sampled HBS light curve presented in the catalog contains 166 data points, while the largest number is over 16,500 (with mean and median values equal to about 5600 and 2700, respectively). For the LMC and SMC, the exposure time of each field was 150 s. There are only a few light curves that contain less than 550 data points, while the maximum number is about 950 (the mean and median values are about 820). To provide color information of the sources, about $10\%$ of observations were taken through a Johnson $V$-band filter with 150~s of exposure time. The typical uncertainty of a single photometric measurement for stars with $I\sim14$ mag and $I\sim18$ mag is approximately 5 mmag and 30 mmag, respectively.
	
	The photometry of the OGLE data was obtained with difference image analysis (e.g., \citealt{1998ApJ...503..325A}; \citealt{2000AcA....50..421W}). For more technical details about processing data from raw images to standard photometric-system light curves, we refer the reader to \cite{2015AcA....65....1U}. The corrections of photometric uncertainties were carried out based on the work of \cite{2016AcA....66....1S}.
	
	In the catalog, we also provide data from earlier phases of the OGLE project -- OGLE-II and OGLE-III, which took place from 1997 to 2000 and from 2001 to 2009, respectively. About $90\%$ of our sample of HBSs were observed during the OGLE-III (the mean number of epochs is about 720) and $30\%$ during the OGLE-II (350 epochs on average). For more information about the observation strategy, data processing, and Warsaw telescope equipment during the OGLE-II, we refer to \cite{1997AcA....47..319U}, and during the OGLE-III we refer to \cite{2003AcA....53..291U}.

	\section{Searching for Heartbeat Stars}
	The presence of ellipsoidal binary systems with high eccentric orbits in the OGLE database was already highlighted in the work of \cite{2004AcA....54..347S}. However, these stars and the ones discovered during subsequent surveys for variable stars in the OGLE data were cataloged as ellipsoidal variables, without any specific remarks. With this in consideration, we began searching for HBSs in the published OGLE catalogs of ellipsoidal variables. In the next step, we decided to inspect the catalogs of eclipsing binaries located in the Magellanic Clouds (MCs) and toward the GB. Finally, we searched for HBSs in the proprietary OGLE data. In the following three subsections, we present the details and results of the searching process. In Table \ref{tab:numbers}, we present the numbers of found HBS candidates in each specific location and catalog.
	
	\subsection{Catalog of Ellipsoidal and Eclipsing Systems} 
	
	The OGLE Collection of Variable Stars (OCVS) contains 450,598 eclipsing binaries found toward the GB (\citealt{2016AcA....66..405S}) and among them 25,405 ellipsoidal variables. The OCVS also contains 48,605 eclipsing binaries located in the MCs (40,204 in the LMC and 8401 in the SMC), which include 1159 and 316 ellipsoidal variables in the LMC and SMC, respectively (\citealt{2016AcA....66..421P}). In order to distinguish HBSs from classical ellipsoidal variables, we visually inspected each light curve from the OGLE catalogs of these stars, looking for any deviations from a standard sinusoidal-like shape. We used the periods provided in the catalog. As a result, we found 486 candidates for HBSs among ellipsoidal variables in the GB. In the sample of ellipsoidal variables from the MCs, we found 34 candidates for HBSs in the LMC and three in the SMC. 
	
	We also looked for HBS candidates in the sample of 1546 ellipsoidal RGs presented by \cite{2004AcA....54..347S}. During the visual inspection of light curves, we selected 112 HBS candidates. After cross-matching this catalog with the one described in the previous paragraph, we found that 38 objects overlap between those catalogs, including four HBS candidates. 
	
	To summarize, out of the total number of 28,353 ellipsoidal variables cataloged in the OCVS (in all locations), we found 631 candidates for HBSs, which is only $2.2\%$ of the whole sample of the ellipsoidal systems. 
	
	In the next stage, we searched for the HBSs among the OGLE eclipsing binaries. We decided to visually inspect the light curves of all eclipsing systems from the MCs and systems with orbital periods longer than 10 days in the GB (a cut was made to limit the size of the sample that will be visually inspected). As a result, we found 22 candidates for HBSs in the GB, 66 in the LMC, and 24 in the SMC. 
	
	\subsection{Catalog of Miscellaneous Objects}
	
	When searching for stars of specific variability type and creating the OCVS, objects with an unknown or unsure type are accordingly marked as \textit{miscellaneous} or \textit{other} type. These objects usually do not appear in the final catalog until their variability type is known with a high confidence level. In total, we searched through more than 13,000 such objects located toward the GB and about 20,000 in the MCs. The miscellaneous stars' catalog may include objects with light curves very similar to the given variability class, but having different parameters (e.g., period, amplitude, or color). On the other hand, the light curves of the HBSs may mimic those of other types of variable stars, such as eclipsing binaries, Be stars, Ap or other spotted stars, or even some pulsating variables; thus, the catalog of miscellaneous objects seemed to be the right place to search for HBSs.
	
	We inspected the whole sample twice, using two different methods of period determination. In the first inspection we used the \texttt{FNPEAKS}\footnote{\url{http://helas.astro.uni.wroc.pl/deliverables.php?active=fnpeaks}} code (created by Z. Ko\l{}aczkowski, W. Hebisch, and G. Kopacki), which calculates the Fourier frequency spectrum for a single-object light curve. In the second inspection we obtained periods using the \texttt{TATRY} code (\citealt{1996ApJ...460L.107S}), based on the analysis of variance method. The period with the highest signal-to-noise ratio (S/N) frequently was equal to half or one-third of the actual value. We noticed that a HBS light curve phase-folded with a wrong period can be easily misclassified, e.g., as a classical ellipsoidal system or spotted star, and thus caution is needed. We found four new candidates for HBSs in the GB, 26 in the LMC, and 13 in the SMC.

	\subsection{Catalog of Nonperiodic and Peculiar Objects}
	The last part of the whole set of HBSs was found during an ongoing project, which is devoted to searching for nonperiodic or peculiar variable stars located in the LMC, using the OGLE-IV data. HBSs often have very sharp and asymmetric light curves, which usually exhibit clear maxima and/or minima. Such properties cause a lot of difficulties in the searching process because typical methods, which rely on Fourier analysis (e.g., \texttt{FNPEAKS} and Lomb-Scargle periodograms) or the ones which have been created to find specific types of variability (e.g., box least-squares periodograms), could show very low S/N in the case of HBSs, what could lead to omitting them during the filtering process.
	
	In the mentioned project, we obtained periods using \texttt{FNPEAKS}, but we did not use any kind of period-related cuts, lowering the risk of omitting HBSs. During this stage of the searching process, we found 205 new candidates for HBSs, which is about half of the whole LMC HBS sample.
	
	\begin{deluxetable}{ccccc}
		
		\tablecaption{Numbers of HBS Candidates Found in the Searching Process}
		\label{tab:numbers}

		\tablehead{\colhead{Location} & \colhead{ELL} & \colhead{ECL} & \colhead{MISC} & \colhead{NP\&PEC} \\ 
			\colhead{} & \colhead{} & \colhead{} & \colhead{} & \colhead{} } 
		
		\startdata
		GB & 486 & 22 & 4 & $-$ \\
		LMC & 142 & 66 & 26 & 205 \\
		SMC & 3 & 24 & 13 & $-$ \\
		GB+LMC+SMC & 631 & 112 & 43 & 205 \\
		\enddata
		\tablecomments{Columns: (Location) The location of the object (GB, LMC, or SMC); (ELL) The OGLE catalogs of ellipsoidal variables; (ECL) The OGLE catalogs of eclipsing binaries; (MISC) The nonpublic OGLE catalog of miscellaneous objects; (NP\&PEC) The denotation of stars found during the OGLE subproject for nonperiodic and peculiar objects.}
			
	\end{deluxetable}
	
	\subsection{Cross-matching with Other Catalogs of Variable Stars}
	
	The whole sample of HBSs from the Kepler project is away from the OGLE observing fields, thus there is no overlap between these catalogs.
	
	To find matches for our HBS collections in other databases of variable stars, we used the \textit{multicone} and \textit{CDS Upload X-Match} options in the TOPCAT\footnote{\url{http://www.star.bris.ac.uk/~mbt/topcat/}} program (\citealt{2005ASPC..347...29T}). We used a 2'' search radius and the best-match output mode. In the SIMBAD database (\citealt{2000A&AS..143....9W}) we found 509 matches for HBSs from the GB and 237 for the MCs. All the matches, but one for the GB sample, turned out to be OGLE ellipsoidal and eclipsing binaries. This one exception also came from the OCVS, but, surprisingly, from the dwarf novae (DN) catalog created by \cite{2015AcA....65..313M}. The considered object is labeled as OGLE-BLG-DN-100 in the DN catalog and as OGLE-BLG-HB-0048 in ours. The misclassification in the DN catalog was caused probably by the low number of epochs available for this star back in 2015.
	
	In the International Variable Star Index (VSX) catalog of variable stars (\citealt{2006SASS...25...47W}), we found 509 matches in the GB. All of them refer to the OCVS. In the MCs, we found three matches, as follows:
	\begin{itemize}
		\item ASASSN-V J054319.59-690954.3 (OGLE-LMC-HB-0392), flagged as a rotating variable star (\citealt{2018MNRAS.477.3145J})
		\item WISE J054601.5-673554 (OGLE-LMC-HB-0404), flagged as a Cepheid star (\citealt{2018ApJS..237...28C})
		\item 2MASS J05474088-7048169 (OGLE-LMC-HB-0412), flagged as a semiregular variable star (\citealt{2019MNRAS.485..961J})
	\end{itemize}
	
	We also cross-matched our catalog with the ASAS-SN Variable Stars Database (\citealt{2014ApJ...788...48S}; \citealt{2018MNRAS.477.3145J}, \citeyear{2019MNRAS.486.1907J}, \citeyear{2019MNRAS.485..961J}), and we identified five new matches for the GB and the same number for MCs. The results are presented in Table~\ref{tab:match}.
	
	\begin{deluxetable}{ccc}[h]
		
		\tablecaption{Results of Cross-matching Our Catalog of HBSs with the ASAS-SN Variable Stars Database}
		\label{tab:match}

		\tablehead{\colhead{OCVS ID} & \colhead{ASAS-SN ID} & \colhead{ASAS-SN} \\ 
			\colhead{} & \colhead{} & \colhead{Variability Type} } 
		
		\startdata
		OGLE-BLG-HB-0082 & J174737.87-232501.7 & Semiregular  \\
		OGLE-BLG-HB-0161 & J175225.44-300702.9 & Red irregular \\
		OGLE-BLG-HB-0164 & J175238.80-293008.2 & Semiregular  \\
		OGLE-BLG-HB-0372 & J180106.60-300247.1 & Rotating      \\
		OGLE-BLG-HB-0379 & J180119.30-285933.8 & Unspecified   \\
		OGLE-LMC-HB-0020 & J045223.82-682153.9 & Semiregular  \\
		OGLE-LMC-HB-0254 & J052624.38-684705.6 & HBS\tablenotemark{a}\\
		OGLE-LMC-HB-0312 & J053226.44-693122.1 & Semiregular  \\
		OGLE-LMC-HB-0340 & J053516.81-690255.7 & EB \\
		OGLE-LMC-HB-0412 & J054741.02-704815.4 & Semiregular \\
		\enddata
		\tablenotetext{a}{Described in \citeauthor{2019MNRAS.489.4705J} (\citeyear{2019MNRAS.489.4705J}, \citeyear{2021arXiv210413925J})}
	\end{deluxetable}
	Our classification of HBSs was based on visual inspection of the phase-folded light curve followed by fitting a theoretical model of the heartbeat variation (discussed in more detail in Section \ref{sec:discussion}). It appears that objects mentioned in this section were misclassified (except for OGLE-LMC-HB-0254), which could be the result of, for instance, an insufficient number of light-curve epochs.
	
	\section{Discussion}\label{sec:discussion}
	
	\begin{figure*}[]
		\centering
		
		\includegraphics[width=1.0\textwidth]{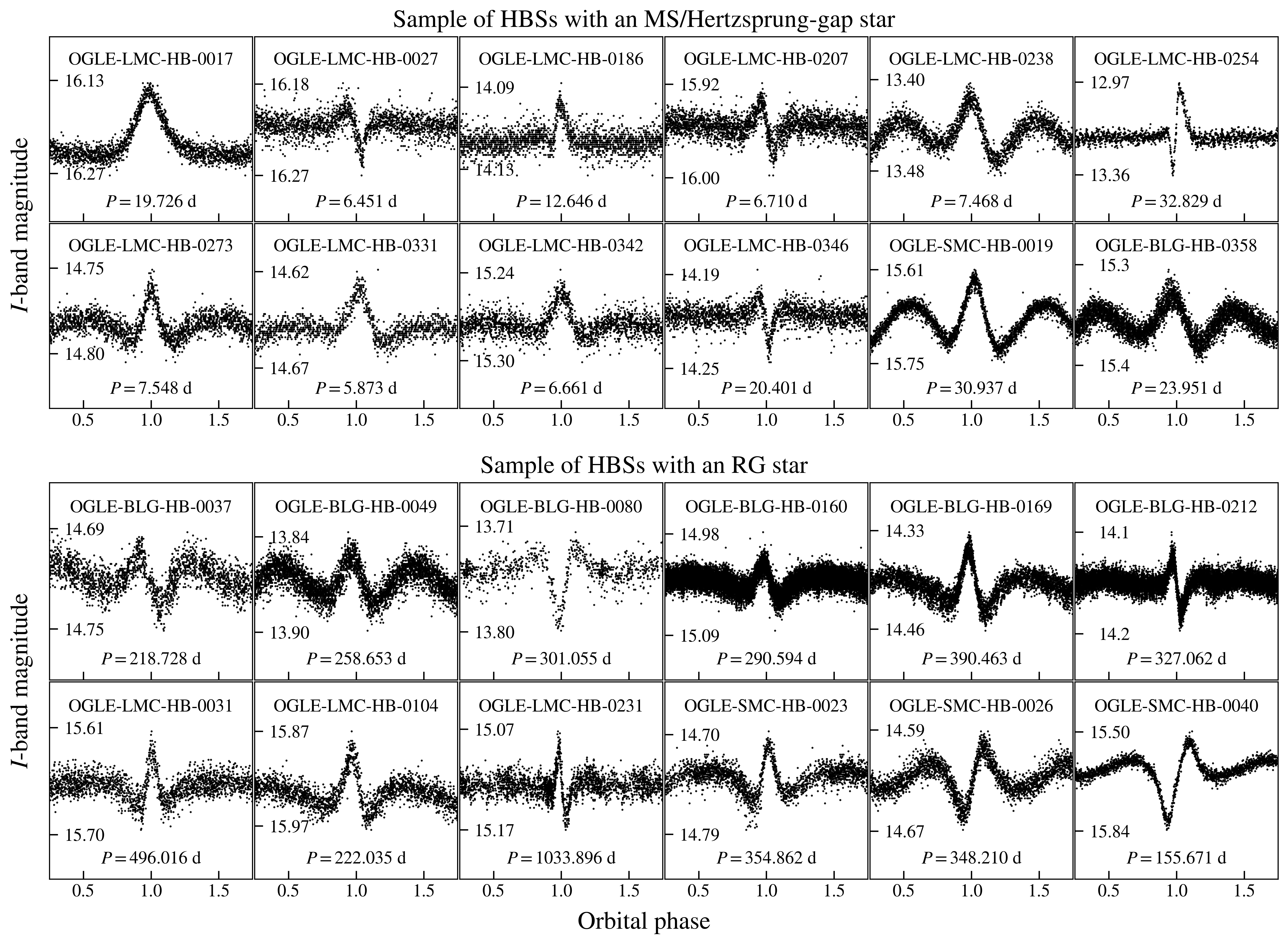}
		\caption{Examples of phase-folded light curves of HBSs from our collection. Here and throughout the paper, the orbital phase equal to 1.0 corresponds to the periastron passage, and $P$ denotes the orbital period.}
		\label{fig:lc_sample}
	\end{figure*}
	
	\begin{figure*}[]
	\centering
	
	\includegraphics[width=1.00\textwidth]{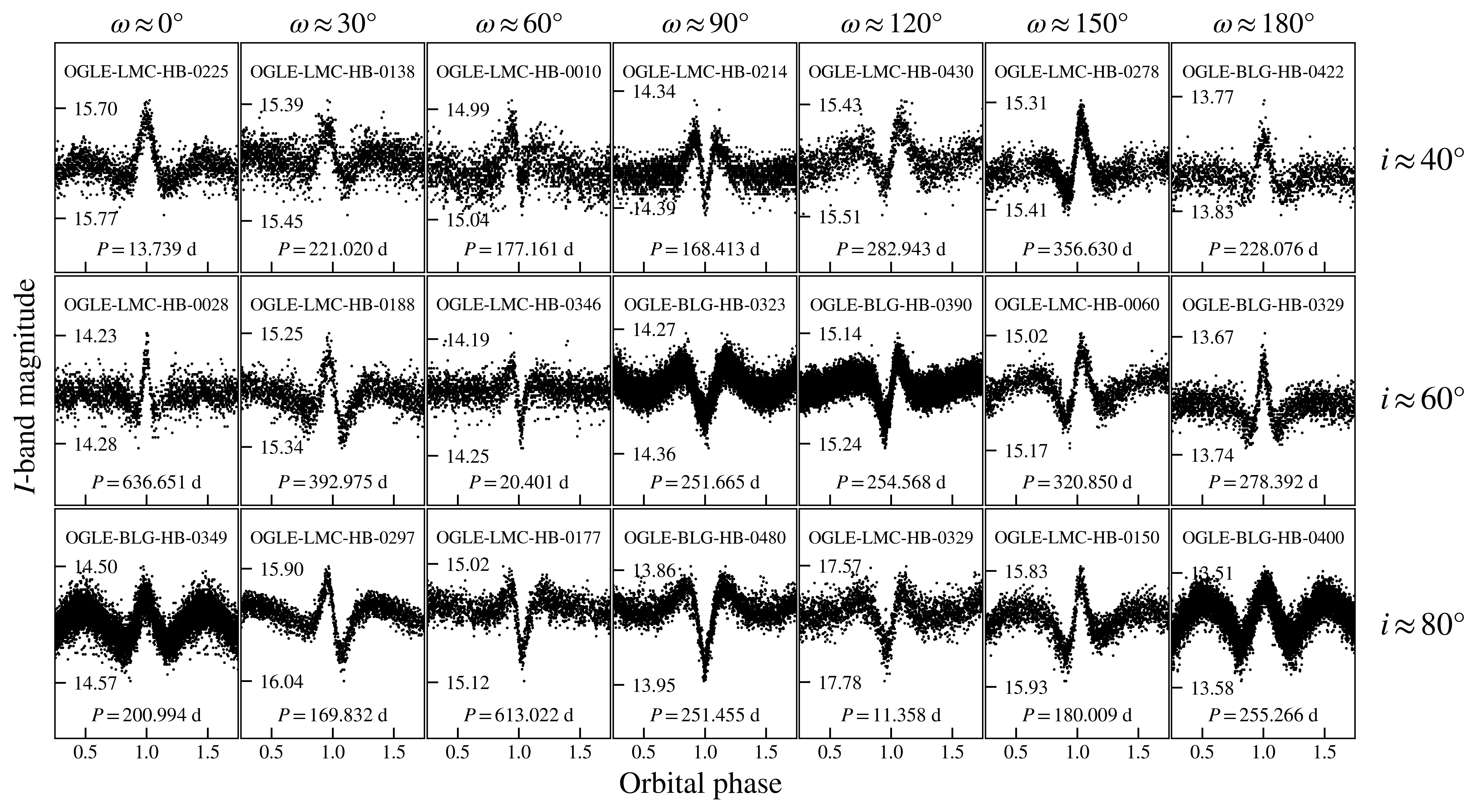}
	\caption{Phase-folded light curves of the sample of OGLE HBSs with different shapes. The orbital inclination, $i$, and the argument of the periastron, $\omega$, were obtained by fitting Kumar et al.'s model to the phase-folded light curves. In the rows, we present HBSs with $i$ equal to circa $40\degree$, $60\degree$, and $80\degree$, respectively. The obtained $\omega$ parameter increases horizontally from $0\degree$ to $180\degree$. The HBSs with $\omega$ equal to about $0\degree$ or $180\degree$, during the heartbeat show two minima and one maximum of brightness, while for $\omega\approx90\degree$ we can see two maxima and one minimum. With increasing $i$ parameter, the minima become deeper. Note, that the right- and left-hand sides in this figure are almost symmetrical reflections of each other.}
	\label{fig:lc_shapes}
	\end{figure*}

	\begin{figure}[]
	\centering
	
	\includegraphics[width=0.48\textwidth]{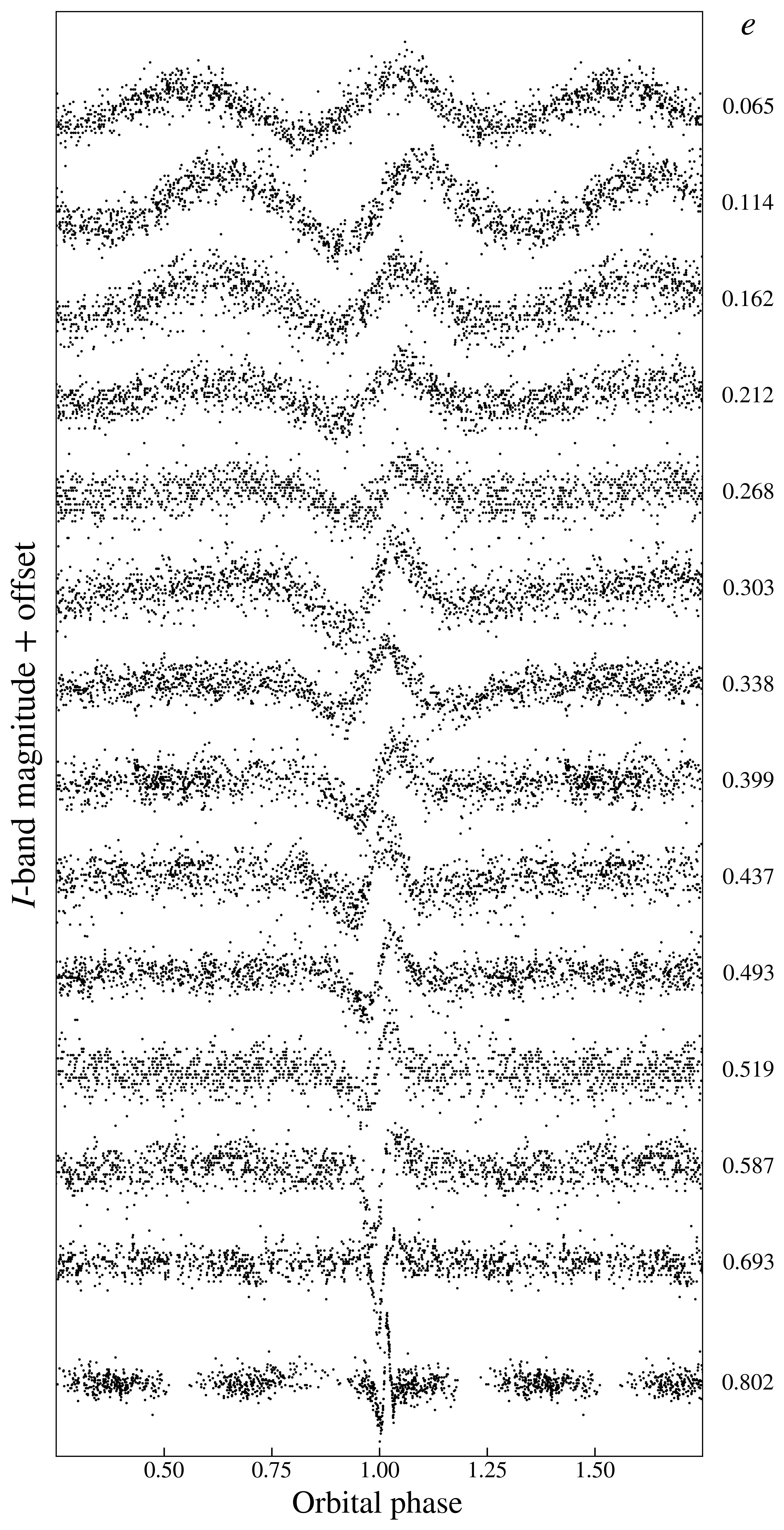}
	\caption{Shape of the HBS light curve as a function of eccentricity, $e$. We used the phase-folded and normalized light curves of the OGLE HBSs. The light curves are shifted in magnitude to avoid overlaps. The $e$ parameter (numbers on the right-hand side) were obtained using Kumar et al.'s model. For low eccentric orbits, the light curves have sinusoidal-like shapes, similar to the classical ellipsoidal variables (top). With increasing $e$, the phase range of the heartbeat becomes narrower (bottom). For $e\approx0.8$, the duration of the heartbeat is about 10\% of the orbital period.}
	\label{fig:lc_ecc}
	\end{figure}
	
	In Figure~\ref{fig:lc_sample}, we present phase-folded light curves for two dozen HBSs from our collection. In the top and bottom panels, we show examples of HBSs which contain a MS/post-MS or RG star as a primary component, respectively. 
	
	To decide whether the object is a potential candidate for a HBS or not, we focused on the shape of the light curve. Each light curve was visually inspected by at least two experienced researchers, who decided if a given star exhibits a heartbeat feature. We excluded each star if at least one of the researchers decided that it is not an HBS candidate. After the initial selection, we performed modeling of the light curves using a convenient analytical model of the flux variations in eccentric binary systems, presented by \cite{1995ApJ...449..294K} (their Equation (44)). Kumar et al.'s model neglects proximity effects other than ellipsoidal variability, such as irradiation/reflection effect and Doppler beaming/boosting. Thus, even a well-fitted model to the light curve derives only an approximation of the orbital parameters. The modeling process of the OGLE HBSs is presented in detail in Section 3 of the related work of \cite{2021arXiv210914614W}. Hereafter, we will refer to this work as Paper II. According to their work, the fractional flux changes ($\delta F / F$) as a function of time ($t$) can be expressed by a corrected version of the Equation (1) presented by \cite{2012ApJ...753...86T}:
	\begin{equation}
		\frac{\delta F}{F}(t)=S\cdot\frac{1-3\sin^2i\sin^2(\varphi(t)+\omega)}{(R(t)/a)^3}+C,
		\label{eq:kumar}
	\end{equation}
	where $C$ is the zero-point offset, $S$ is the scaling factor of the amplitude, $\omega$ is the argument of the periastron, $i$ is the inclination angle of the orbit, $\varphi(t)$ represents the true anomaly as a function of time, $R(t)$ describes the distance between components of the system as a function of time, and $a$ is the semimajor axis.
	
	Kumar et al.'s model of brightness variations predicts a large variety of the possible light-curve shapes. Synthetic models of the light curves generated from the model are shown in Figure 5 in the paper of \cite{2012ApJ...753...86T}. The authors presented a grid of the heartbeats depending on $e$, $i$, and $\omega$ parameters. The~wealth of the OGLE collection of HBSs allows us to create a similar figure, but for the actual light curves. We present these light curves in Figure~\ref{fig:lc_shapes}. In each plot, we show a phase-folded light curve with an orbital period,~$P$. Each row contains HBSs with $i$ equal to circa $40\degree$, $60\degree$, and $80\degree$, from top to bottom. The $\omega$ parameter increases horizontally, from $0\degree$ to $180\degree$. The range of $\omega$ is only a half of the full angle because the stellar distortion due to tidal deformation is symmetric along the elongation axis. Note, that the right and left halves of Figure \ref{fig:lc_shapes} are symmetric reflections of each other.

	\begin{figure*}[]
	\centering
	\includegraphics[width=0.485\textwidth]{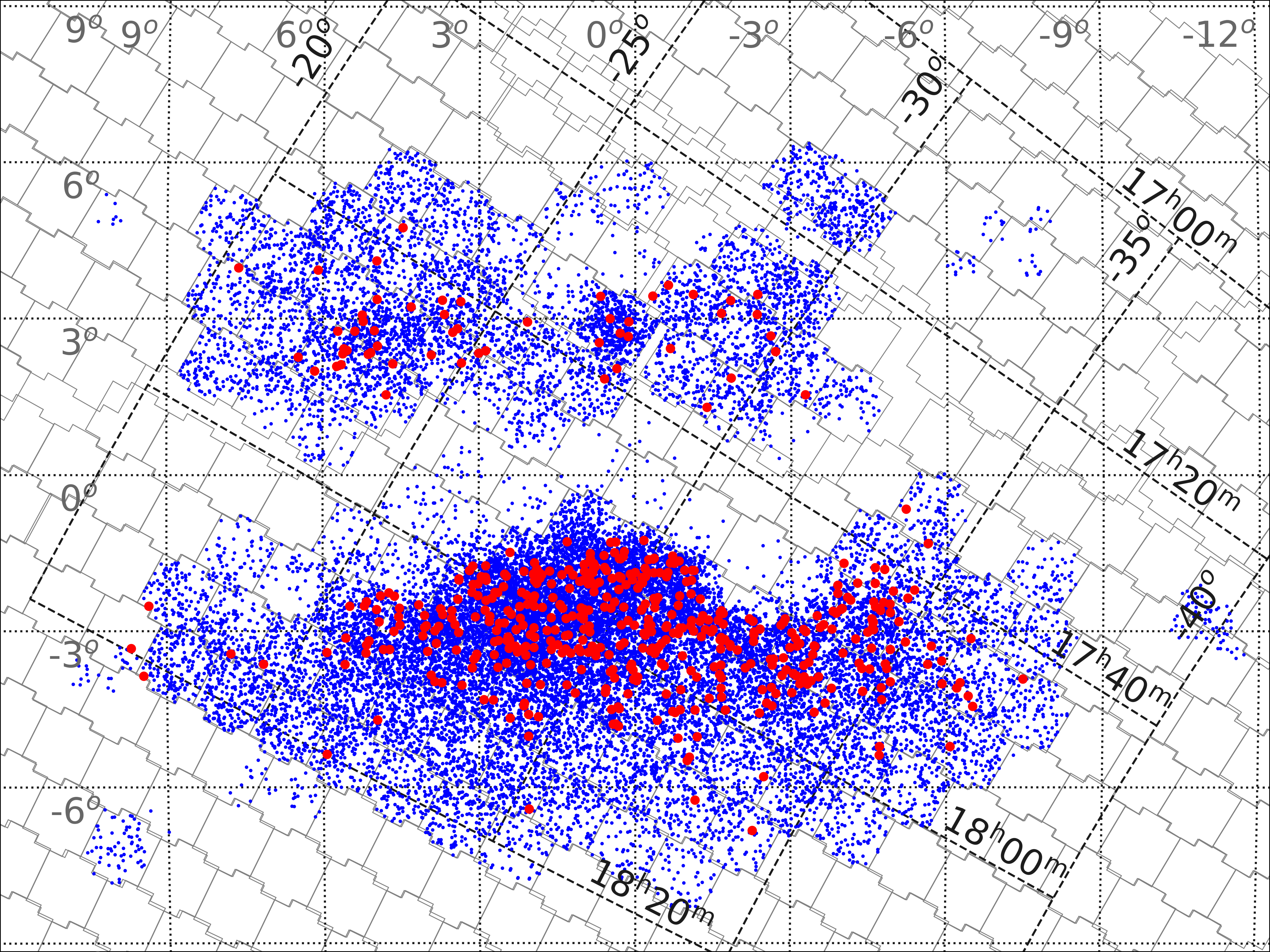} \hspace{0.3cm}
	\includegraphics[width=0.465\textwidth]{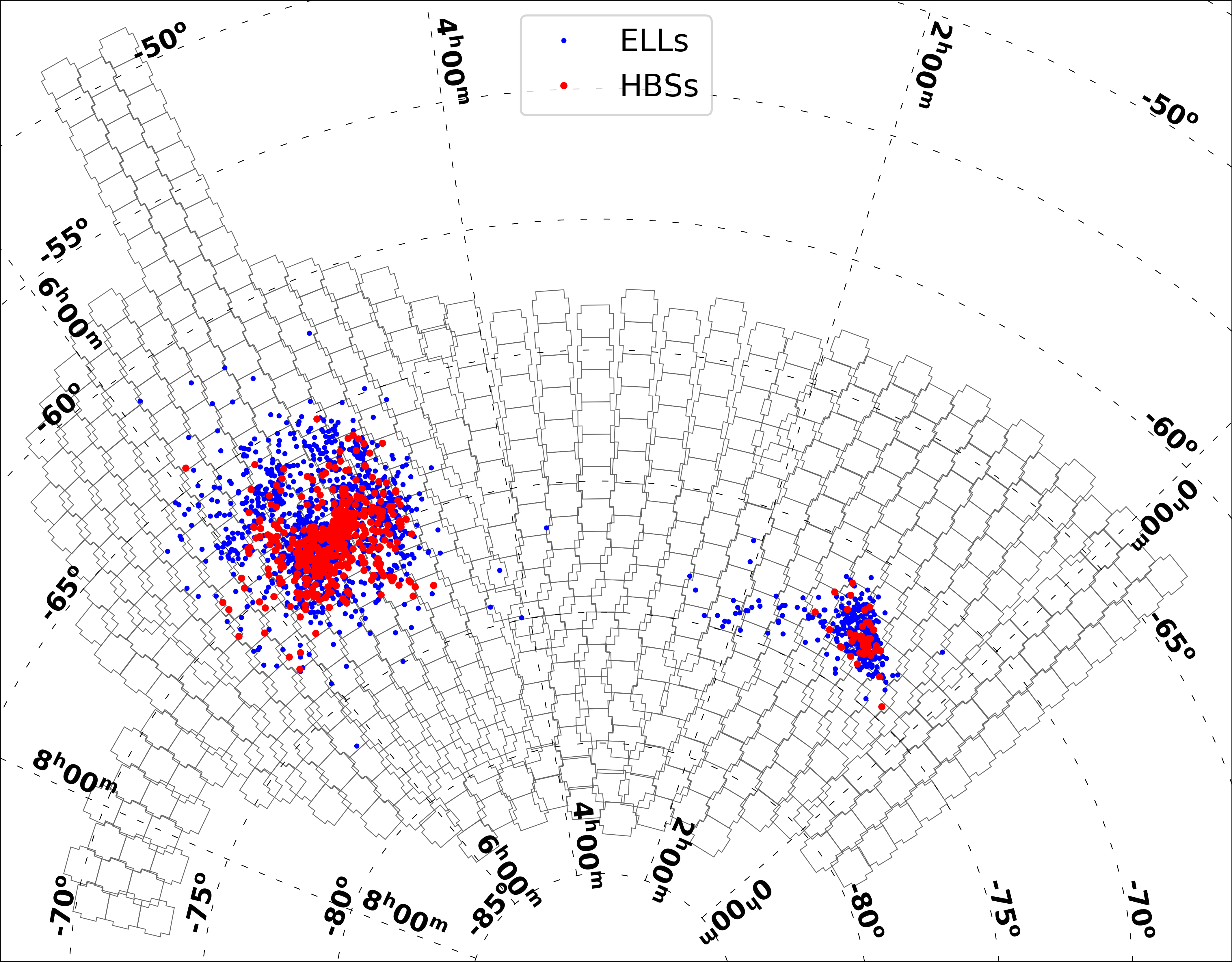}
	
	\caption{Distribution of the HBSs in the sky toward the GB (left panel) and in the MCs (right panel). Black contours correspond to the OGLE-IV footprint. Dotted and dashed black lines denote the galactic and equatorial coordinates, respectively. The blue and red points represent the OGLE ellipsoidal variables and HBSs, respectively. Note that the HBSs from the LMC align along the central bar.}
	\label{fig:maps}
	\end{figure*}

	\begin{figure*}[]
	\centering
	\includegraphics[width=0.9\textwidth]{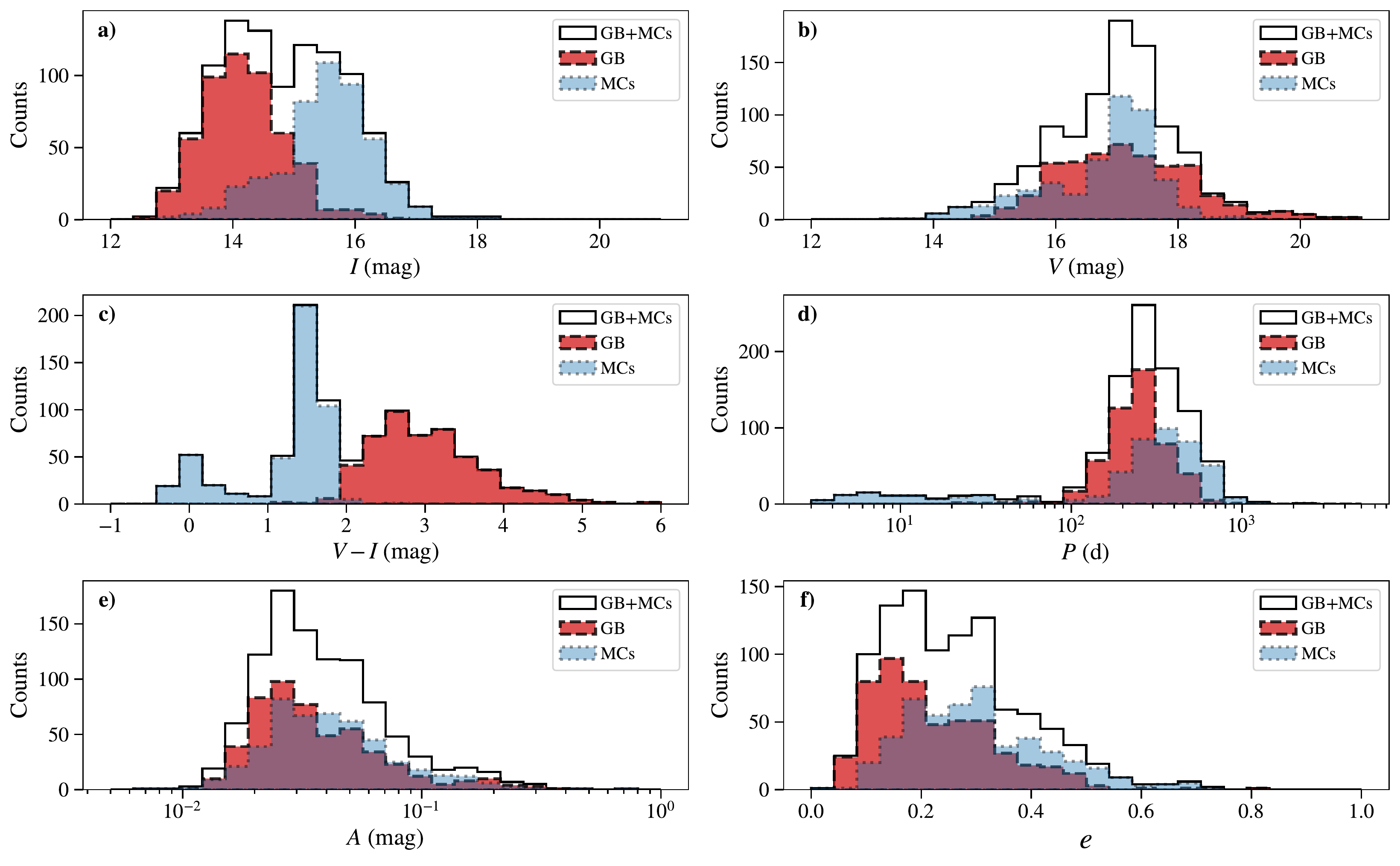}
	\caption{Histograms of basic observational and orbital parameters of the HBSs located toward the GB (red boxes) and in the MCs (blue boxes). With the solid black line, we denote a histogram of the combined samples of the HBSs from all locations. In the panels, we present histograms of: \textbf{a)} the mean $I$-band~magnitude; \textbf{b)} the mean $V$-band~magnitude; \textbf{c)} $V-I$ color index; \textbf{d)}~orbital period, $P$; \textbf{e)} $I$-band peak-to-peak amplitude (based on the fitted model of Kumar et al.), $A$; \textbf{f)} orbital eccentricity, $e$.}
	\label{fig:catalog_hist}
	\end{figure*}

	While the shape of the heartbeat depends mainly on the $i$ and $\omega$ parameters, its duration relative to the orbital period is strongly anticorrelated with the eccentricity. In Figure~\ref{fig:lc_ecc}, we present normalized (in the brightness amplitude) light curves of the HBSs with different eccentricities, which increase downwards. For low eccentric orbits ($e\lesssim0.1$), the light curve has a nearly sinusoidal shape, similar to the classical ellipsoidal variables. With increasing eccentricity, the phase range of the heartbeat gets narrower. For $e\gtrsim0.4$, the light curve starts to exhibit two distinct phases: a narrow heartbeat and a phase of constant brightness. For $e\approx0.4$, the duration of the heartbeat is about $40\%$ of the orbital period, while for $e\approx0.8$, the heartbeat durations decrease to about $10\%$ of the orbital period.

	\subsection{Sample of the HBSs in the Galactic Bulge}
	
	The sample of HBS candidates located toward the GB consists of 509 stars previously cataloged in the OCVS and four stars, which are new discoveries. The position of the GB sample of the HBSs in the sky is presented in the left panel of Figure~\ref{fig:maps}. This sample is dominated by stars lying on or between the red-giant branch (RGB) and the asymptotic giant branch (AGB).
	
	Our sample, contrary to HBSs from the Kepler catalog, consists of long-period systems (mostly a few hundred days) with high-amplitude brightness variations (a few hundredths of a~magnitude and larger). In Figure~\ref{fig:catalog_hist}, for a better picture in the observational context, we present histograms of the basic observational and orbital parameters. These parameters are as follows: $I$- and $V$-band mean~magnitudes, $V-I$ color indices, orbital periods and eccentricities, and Kumar et al.'s model peak-to-peak $I$-band amplitudes expressed in~magnitudes. We can see that the peak of the $I$-band mean~magnitude distribution (panel~\textbf{a)}) for the GB sample of HBSs is about 2~mag brighter than the MCs sample. If we assume the distance to the Galactic center is about $8$ kpc (\citealt{2015ApJ...811..113P}) and to the LMC is about $50$ kpc (\citealt{2019Natur.567..200P}), we get a distance modulus difference of about $4$~mag. The GB and MCs HBSs are dominated by systems consisting of RG stars, thus the distribution of the mean brightness in a given filter should be shifted by that distance modulus difference. This 2~mag difference between the observed peak of the distribution and the distance modulus is the result of high interstellar extinction toward the GB.
	
	In the $V$ band, the impact of the extinction is higher than in the $I$ band, thus the peak of the distribution of the mean $V$-band~magnitude (panel~\textbf{b)}) in the GB is similar to the MCs. The effect of heterogeneity of the extinction depending on the direction toward the GB and different distances between HBSs explain the high scatter of the $V-I$ color index (panel~\textbf{c)}) for the GB sample.
	
	In contrast to the MCs, the GB sample of HBSs includes almost exclusively long-period binaries, which is shown in panel~\textbf{d)} of Figure~\ref{fig:catalog_hist}. The $I$-band amplitudes of brightness variations (panel~\textbf{e)}) are similar to the GB and MCs, and they mostly spread in the range between $0.001$ and $0.01$~mag. For the GB sample of HBSs, however, the eccentricity distribution (panel~\textbf{f)}) is slightly shifted toward lower values and it has a lower scatter. The peak of the eccentricity distribution is about $e\approx0.15$ and it spreads even to $e\approx0.8$.

	\subsection{Sample of the HBSs in the Magellanic Clouds}
	
	Our collection of HBS candidates from the LMC and SMC is much richer in new discoveries than the one from the GB. Out of 479 HBSs, a sample of 237 objects has been cataloged in the SIMBAD database, and in the majority overlaps with the sample of 235 stars cataloged in the OCVS. In total, 190 HBS candidates (189 from the LMC and one from the SMC) are new discoveries. 
	
	From the analysis of the color--magnitude and Hertzsprung--Russell diagrams presented in Paper II (Figures $9-10$ therein), we conclude that the MCs sample of the HBSs can be divided into at least two large groups. The first one consists of hot MS or Hertzsprung-gap stars. This group is also noticeable in panel \textbf{c)} in Figure~\ref{fig:catalog_hist} as a little "hill" near $V-I=0$~mag and \textbf{d)} as bins for period less than 100 days (see Paper II, their Figure 11). The second (much more numerous) group consists of systems with an RG star, which is similar to the results for the GB. 
	
	\begin{figure}[]
		\centering

		\includegraphics[width=0.48\textwidth]{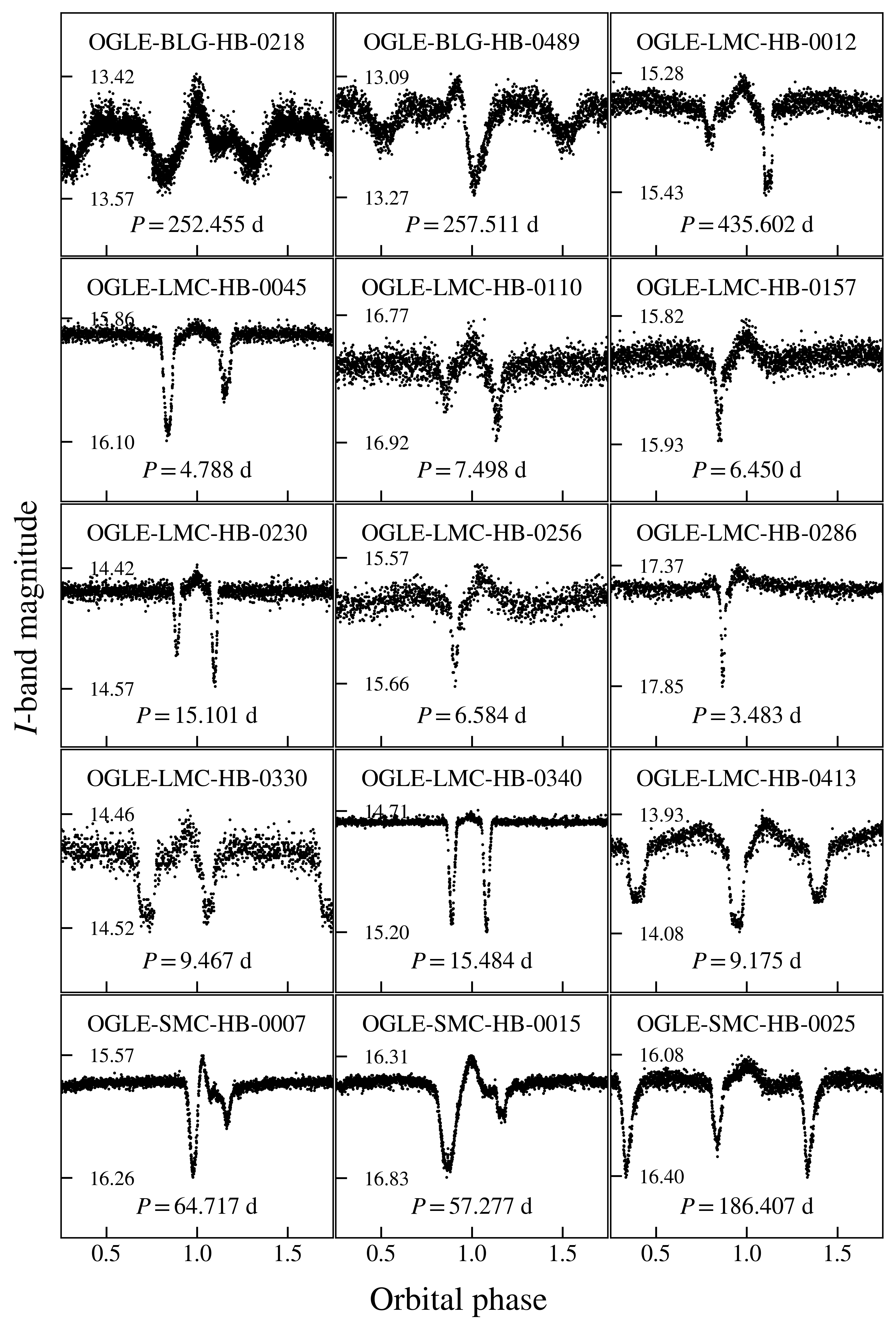}

		\caption{Examples of phase-folded light curves of the eclipsing HBSs from our collection.}
		\label{fig:lc_sample_ecl}
	\end{figure}
	
	\begin{figure}[]
		\centering

		\includegraphics[width=0.48\textwidth]{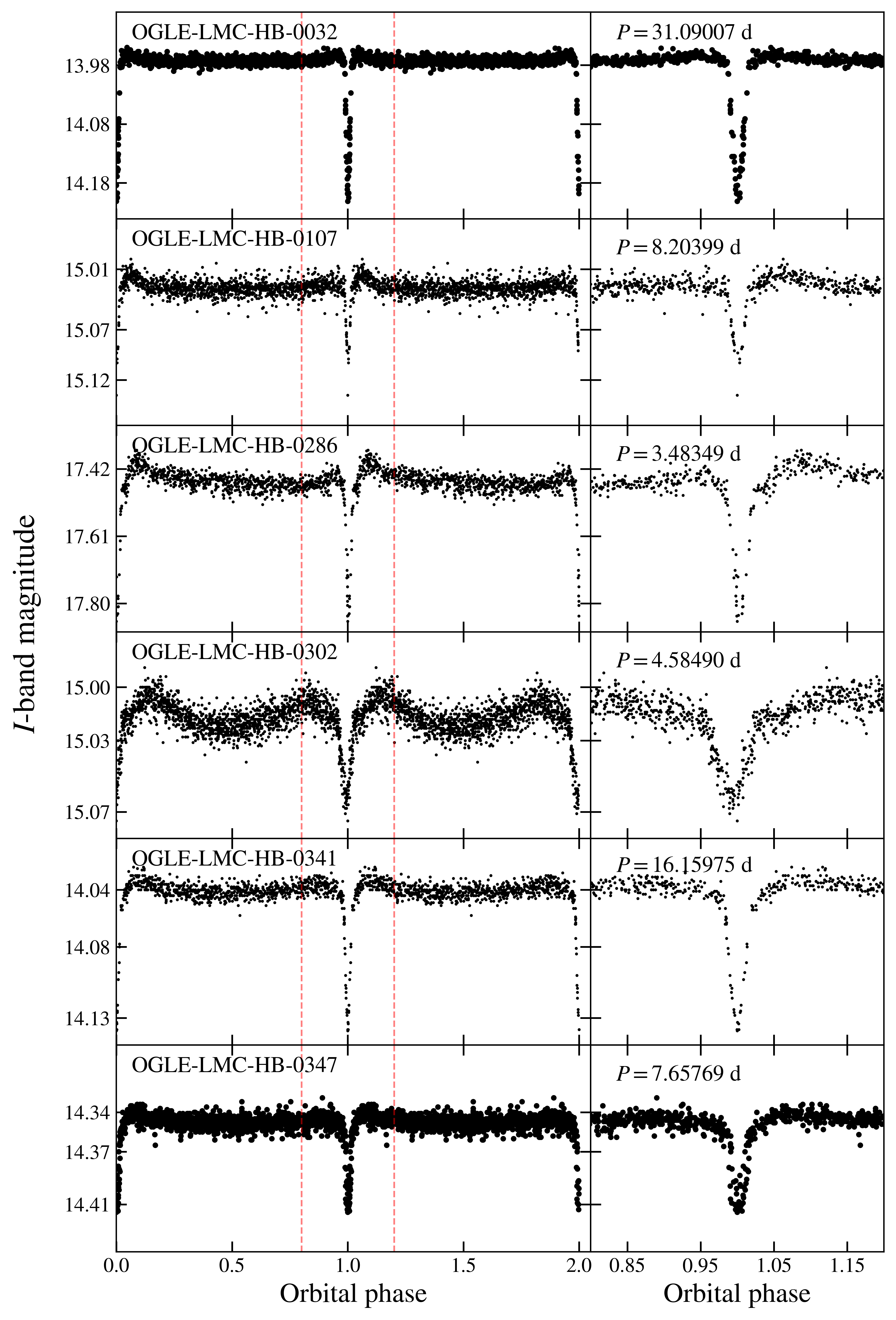}

		\caption{Phase-folded light curves (left panels) and a close-up view on the heartbeat (part of the phase-folded light curve between dashed red lines; right panels) of six eclipsing HBSs, where the eclipse appears very close to the minimum brightness of the heartbeat.}
		\label{fig:lc_misc2}
	\end{figure}

	\begin{figure}[]
		\centering

		\includegraphics[width=0.48\textwidth]{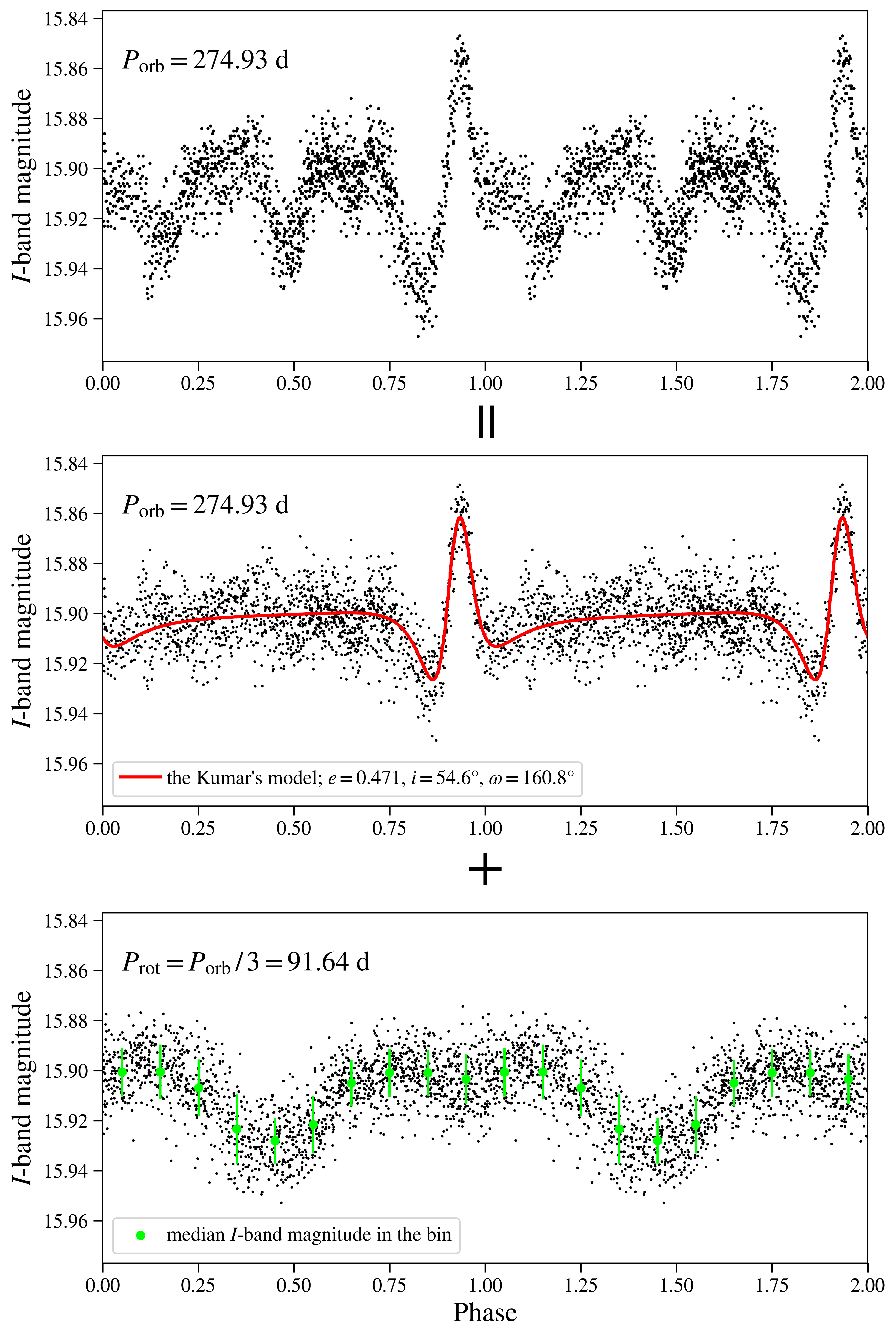}
		
		\caption{Phase-folded light curve of the OGLE-LMC-HB-0287 (top panel). The light curve is a combination of the heartbeat variation with an orbital period, $P_{\rm orb} = 274.93$ days (middle panel) and dimmings, which are probably caused by a dark spot on the primary's surface, with a rotational period, $P_{\rm rot}=P_{\rm orb}/3= 91.64$ days (bottom panel). In the middle panel, we present the light curve from the top panel after subtracting the dimming caused by the spot. With the red line we plot the fitted Kumar et al.'s model. In the bottom panel, we show a phase-folded light curve with $P_{\rm rot}$ after subtracting the heartbeat variation. With green dots we mark the median $I-$band magnitude in the 0.1 phase bins. Green bars represent a standard deviation of the magnitude in the bin.}
		\label{fig:spotted}
	\end{figure}
	
	\subsection{HBS with Additional Variations}
	Both light curves of ellipsoidal variables with circular orbits and of HBSs may be affected by additional variations, including intrinsic changes, such as spots or stellar pulsations, and extrinsic events, e.g., eclipses. Noticeable additional variations are present in about 380 HBSs. In many cases, these changes hampered or even prevented finding a proper Kumar et al.'s model during the fitting process. Such troublesome HBSs have been accordingly flagged in the last column of Table \ref{tab:hb}. Orbital parameters for models marked with a zero flag should be used with caution. These parameters may not reflect the true values, so their use in statistical analyses may lead to biased or false conclusions.
	
	\begin{figure*}[]
		\centering
		
		\includegraphics[width=1.0\textwidth]{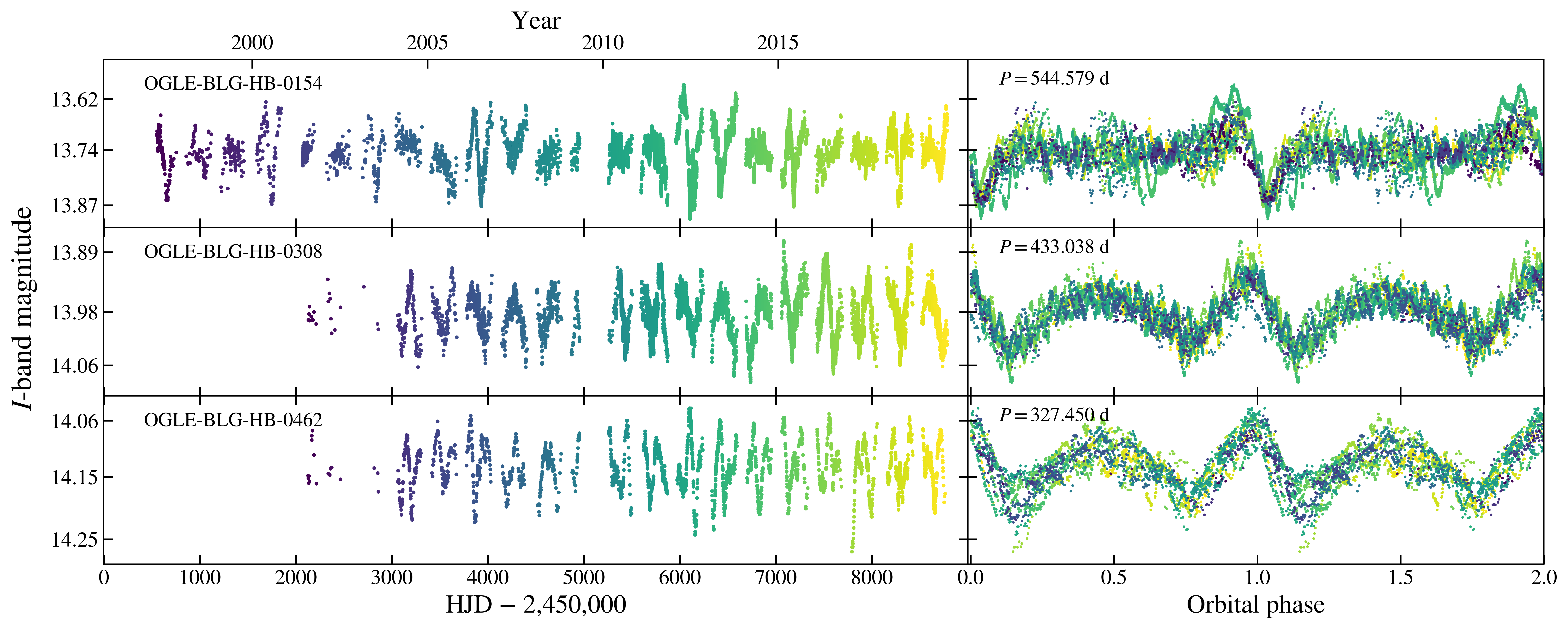}

		\caption{Examples of time-domain (left panels) and phase-folded light curves (right panels) of HBSs exhibiting OSARG pulsations. Colors of the points denote the observation time. The color scale is the same in all panels.}
		\label{fig:lc_osarg}
	\end{figure*}
	
	\begin{figure*}[]
		\centering

		\includegraphics[width=1.0\textwidth]{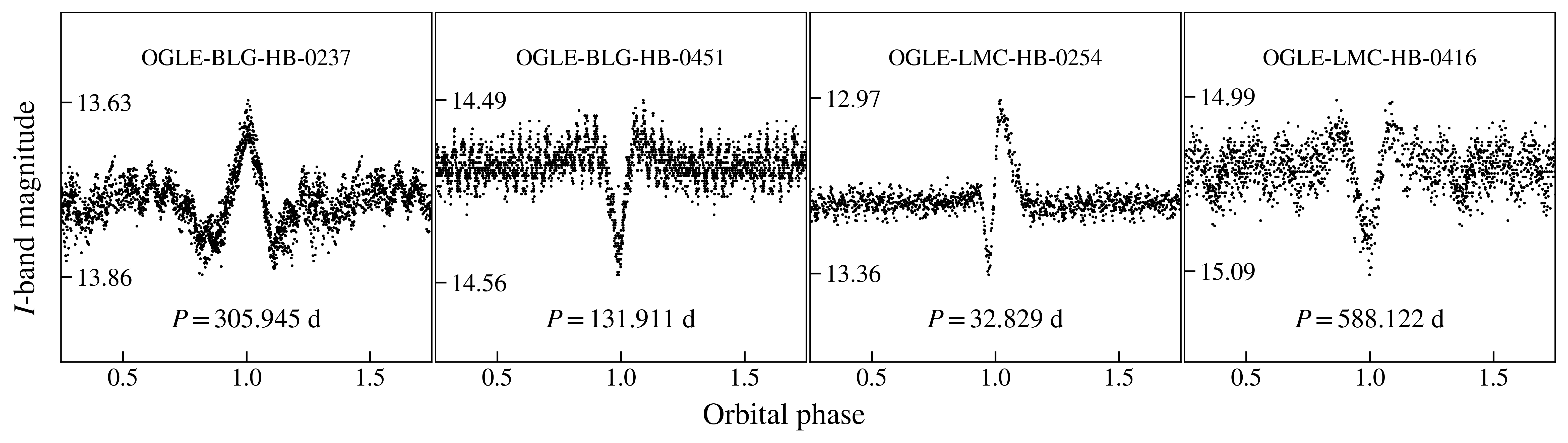}

		\caption{Phase-folded light curves of the sample of HBSs exhibiting high-amplitude TEOs. The TEOs are visible as oscillations outside periastron.}
		\label{fig:lc_teo}
	\end{figure*}
	
	\subsubsection{Eclipses or Spots}
	In about 140 objects, besides heartbeat variation, we observe additional dimmings in the systems' brightnesses. For most stars, these dimmings are most probably the results of eclipses. However, in some cases, such dimmings could be the result of a spot on the star's surface (e.g., \citealt{2019ApJ...879..114I}). HBSs with noticeable eclipses are also present in the catalog of \cite{2016AJ....151...68K}. Heartbeat variation in the eclipsing system was also noticed recently by  \cite{2021MNRAS.504.3749P} and \cite{2021arXiv210606329K}.
	
	The probability of an eclipse rises with a decreasing distance between stars, and so increasing chances to catch the eclipse near the periastron passage, even if the orbital inclination is not near $90\degree$. However, to observe an eclipse, the orbit should be properly oriented to the observer. Simulations of the heartbeat variation show that dimmings in the light curve occur when the tidal bulge is oriented toward or outward relative to the observer. Written differently, for a HBS, the minima of the light curve appear near the superior and inferior conjunction. Therefore, if the inclination is high enough that the components form an eclipsing system, at least one of the eclipses will always occur near the minimum of the heartbeat, while the second eclipse may occur in any orbital phase.
	
	In Figure~\ref{fig:lc_sample_ecl}, we present light curves of 15 HBSs in eclipsing systems. We can see that for many systems both eclipses appear near the heartbeat, which occurs near the periastron passage (e.g., OGLE-LMC-HB-0045, 0110, 0230, 0340, OGLE-SMC-HB-0007, 0015), but there are also objects for which one of the eclipses appears at different orbital phase (e.g., OGLE-BLG-HB-0489, OGLE-LMC-HB-0330). For a subsample of HBSs, we can see both primary and secondary eclipses (e.g., OGLE-BLG-HB-0218, OGLE-LMC-HB-0012, 0413, OGLE-SMC-HB-0025), while for others only the primary ones (e.g., OGLE-LMC-HB-0157, 0256, 0286). In Figure~\ref{fig:lc_misc2}, we show a close-up view on the light curves of six systems, where we can see only one eclipse near the heartbeat's minimum.
	
	In the object assigned as OGLE-LMC-HB-0287, we noticed dimmings which can hardly be explained by eclipses. We present the phase-folded light curve of this object in the top panel of Figure \ref{fig:spotted}. During one orbital phase we can see three distinct minima and two of them occur outside the heartbeat. One of the possible explanation of such variability is the presence of a dark spot on the primary's surface. In the middle panel of Figure \ref{fig:spotted}, we present a phase-folded light curve with orbital period, $P_{\rm orb}$, after subtracting dimmings caused by the spot. Assuming a pseudo-synchronous rotation of the primary with an one-third orbital period, the observer will see three dimmings caused by the spot and a heartbeat variation near the periastron passage. The observed light curve is then a combination of these two effects. With the red line, we plot the fitted Kumar et al.'s model. In the bottom panel we preset a phase-folded light curve with rotation period, $P_{\rm rot}=P_{\rm orb}/3$ without the heartbeat variation.

	\subsubsection{OSARGs}
	OGLE small-amplitude red giants (OSARGs) are a subclass of long-period variables. They are RGB or AGB stars that show multi-periodic light variations with periods from 10 to about 100 days (e.g., \citealt{2004MNRAS.349.1059W}, \citealt{2004AcA....54..129S}). As shown in Paper II, in the period--luminosity diagram (their Figure 12), HBSs occupy the space near the long secondary period and ellipsoidal variables, which often exhibit OSARG-like pulsations; thus, this kind of variability among the HBSs is not unexpected. The presence of OSARG-like oscillations among HBSs is also valuable for asteroseismology, because it confirms that the periodic tidal force does not completely suppress intrinsic oscillations. In Figure~\ref{fig:lc_osarg}, we present three HBSs with clearly visible OSARG variations. We noticed such changes in 176 HBSs ($\approx20\%$ of the RG HBSs).
	
	\begin{figure*}[]
		\centering

		\includegraphics[width=1.0\textwidth]{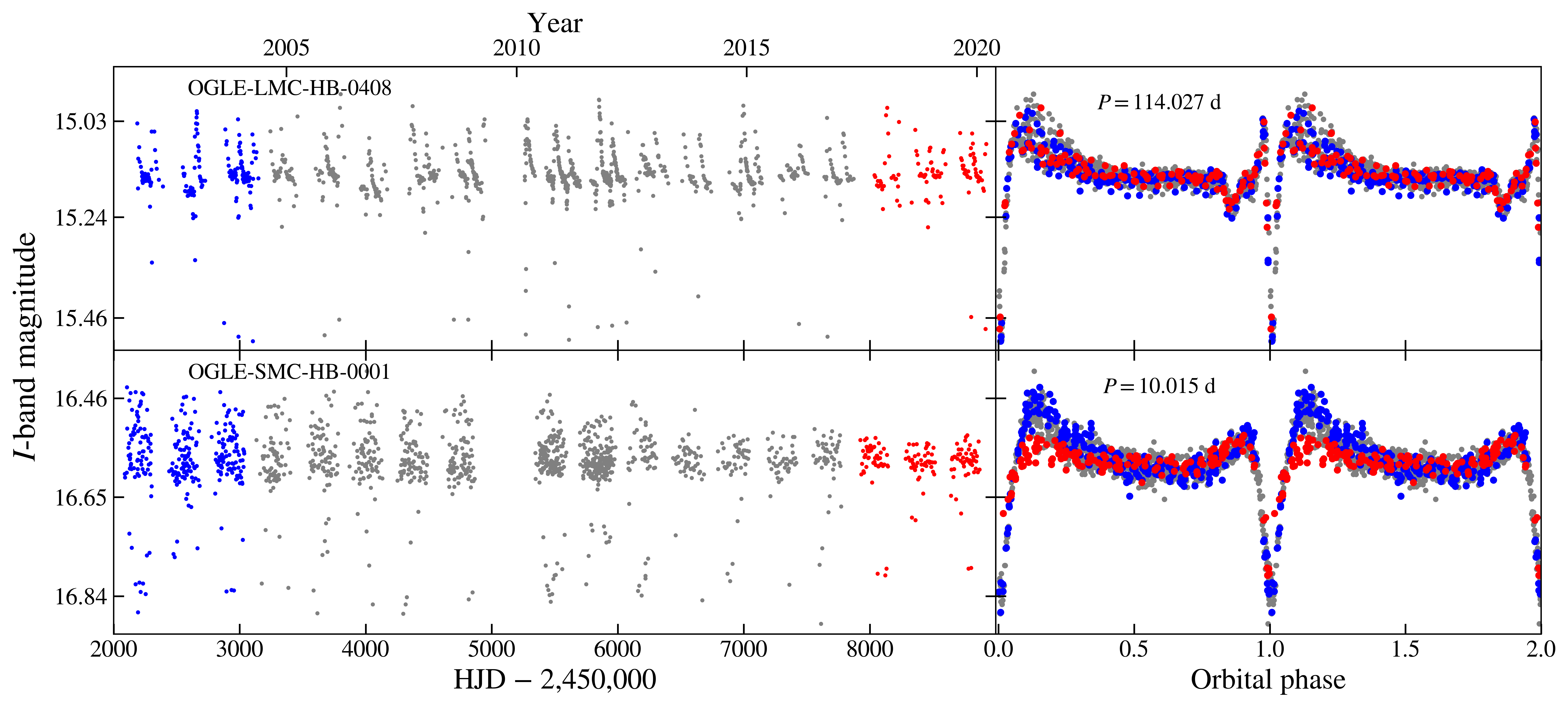}

		\caption{Light curves of two HBSs showing unusual brightness variations. Time-domain (left panel) and phase-folded light curves (right panel) are presented. We marked the first and last three observational seasons with red and blue colors, respectively.}
		\label{fig:lc_misc1}
	\end{figure*}
	
	\subsubsection{Tidally Excited Oscillations}
	The gravitational interactions between components of a binary system can induce tidally excited oscillations (TEOs; e.g., \citealt{1975A&A....41..329Z}, \citealt{1980A&A....92..167H}, \citealt{1998ApJ...499..853E}, \citealt{1995ApJ...449..294K}, \citealt{2017MNRAS.472.1538F}). These oscillations, contrary to self-excited pulsations of the star, appear at frequencies exactly equal to integer multiples of the orbital frequency, therefore they phase well with the orbital period. In our sample of the HBSs, we found 52 systems exhibiting TEOs (see Paper II, their Section 5), which is 5\% of the whole sample. This percentage, however, is the lower limit of HBSs exhibiting TEOs, because the detection threshold in the Fourier frequency spectra is between 0.2 and 3 mmag. In Figure~\ref{fig:lc_teo}, we present phase-folded light curves of four OGLE HBSs exhibiting noticeable TEOs. They are visible as pulsations outside the heartbeat phase.
	
	\subsubsection{Miscellaneous Variations}
	In two HBSs, designated as OGLE-LMC-HB-0408 and OGLE-SMC-HB-0001, we detected unusual changes which occur right after the heartbeat variation. We present the light curves of these stars in Figure~\ref{fig:lc_misc1}. In the left panel, we show the entire light curves, and in the right panel, the phase-folded ones. 
	
	To highlight the time evolution of these atypical changes, we marked the data from the first three and the last three seasons ($\sim$20 yr time span) with different colors. We can see that in OGLE-LMC-HB-0408 these changes are rather stable for the entire time span, while in OGLE-SMC-HB-0001 they almost completely disappeared. In the light curve of OGLE-SMC-HB-0001, besides the change of the heartbeat shape, we also see a slight decrease in the depth of the eclipse. 
	
	The change of the shape of OGLE-SMC-HB-0001's light curve could be a result of a high apsidal motion rate. A rotation of the line of apsides can be observed in a system consisting of two components (e.g., \citealt{2014A&A...572A..71Z}), as well as in a binary system with a tertiary component (e.g., among eclipsing binaries, \citealt{2015AJ....149..197Z}; and in HBS, KIC 3749404, \citealt{2016MNRAS.463.1199H}). Since there is no noticeable change in the shape of the OGLE-LMC-HB-0408 light curve during all observational seasons, the unusual changes of brightness are unlikely caused by apsidal motion. A more likely explanation is a mass transfer between components because the unusual changes happen right after the heartbeat, which occur near the periastron passage.
	
	\section{The Catalog Content and Data Availability}
	The OGLE collection of HBSs found toward the GB and MCs contains 512 and 479 candidates (439 in the LMC and 40 in the SMC), respectively. The data for all these stars are available at the OGLE websites:
	\begin{itemize}
		\item \url{https://www.astrouw.edu.pl/ogle/ogle4/OCVS/blg/hb/} for the GB.
		\item \url{https://www.astrouw.edu.pl/ogle/ogle4/OCVS/lmc/hb/} for the LMC.
		\item \url{https://www.astrouw.edu.pl/ogle/ogle4/OCVS/smc/hb/} for the SMC.
	\end{itemize}
	
	Each object has an individual identifier according to the location (GB, LMC, SMC): OGLE-BLG-HB-NNNN, OGLE-LMC-HB-NNNN, OGLE-SMC-HB-NNNN (where NNNN is a four-digit number). The numbers are assigned in an ascending R.A. order. In Table \ref{tab:ident}, we present the sky coordinates, the OGLE database identifiers, and also the cross-matched IDs with OCVS and (if not found) with other surveys. To simplify the finding of an object in the sky, we also provide $1'\times1'$ finding charts cropped from OGLE template images, which are available at the OGLE websites. In Table \ref{tab:hb}, we collected all basic observational information and derived orbital parameters from the light-curve modeling using an analytical model of brightness variations (\citealt{1995ApJ...449..294K}). In Table \ref{tab:kumar}, we present the derived parameters and their uncertainties from the light-curve modeling. To fit Kumar et al.'s model to the $I$-band light curve, we used a Markov Chain Monte Carlo (MCMC) method. For more details about the modeling process, we encourage the reader to acquaint themselves with Paper II. 
	
	The catalog also includes all available $I$- and $V$-band time-series photometry obtained during the OGLE-II, OGLE-III, and OGLE-IV surveys. The data for both passbands and all of the OGLE phases were separately calibrated to the standard Cousins $I$- and Johnson $V$-photometric systems. For each light curve obvious outliers were removed. For individual stars, the light curves in different phases of the project may be shifted in magnitude. The offset between light curves is most probably the result of different instrumental configurations of filters and CCD detectors or problems with the DIA photometry pipeline. This should be taken into consideration in the light-curve merging process (if needed).

	\section{Conclusions and Future Work}
	We presented a collection of almost one thousand HBS candidates found in the OGLE project database. The collection includes 512 objects located toward the GB, 439 in the LMC, and 40 in the SMC. Our sample of HBSs consists of objects of various spectral types and luminosity classes. 
	
	The largest group contains evolved binaries with a primary star located on or between the RGB and AGB. The second numerous group consists of systems in an early evolutionary stage, where the primary star is mostly a hot MS/post-MS star.
	
	We provide time-series photometry in the $I$- and $V$-bandpass obtained during the OGLE-II, OGLE-III, and OGLE-IV projects (up to 25 yr of coverage), a finding chart for each object, and numerous information about individual objects, such as coordinates, orbital period, $I$-band amplitude, mean brightness, eccentricity, orbital inclination, and argument of periastron. All the data are available on the OGLE website.
	
	In Paper II, we described in more detail the physical properties of our sample of the HBSs, mainly derived from the light-curve modeling using Kumar et al.'s model.
	
	The presented catalog increases the number of known HBSs fivefold. The existence of binary systems with eccentric orbits and containing a massive MS star, on the one hand, and less massive but more evolved primary, on the other hand, challenges theories of the binaries' evolution, with emphasis on the orbit circularization mechanisms.
	
	Our sample may be also used as a testbed for a detailed study of TEOs, especially using high-cadence time-series photometry, provided, for instance by TESS, at present, or by The Nancy Grace Roman Space Telescope (formerly known as WFIRST), in the future.
	
	Last but not least, the OGLE HBSs are valuable targets for spectroscopic observations. First, the radial-velocity curves will allow for confirmation of the binary nature of candidates. Second, high-resolution spectra will lead to a better estimation of the physical parameters of the primary and may also provide information about the secondary component. Third, the simultaneous modeling of radial-velocity changes and light curves will allow for both independent determination of orbital parameters and verification of the correctness and limitations of Kumar et al.'s model. Fourth, the time-series of spectra may be used to verify if any varying emission is present due to, e.g., the surrounding disk, periastron Roche-lobe overflow, colliding or tidally enhanced stellar winds.
	
	In the future, we plan to extend our catalog for new candidates. We seek to find them mainly during the search for peculiar variable stars toward the MCs, which is a part of one of the OGLE subprojects. We assess to find a few hundred such objects.

	\section*{Acknowledgment}
	
	We are grateful to the anonymous referee for many inspiring comments that helped to improve this manuscript. We would like to thank Prof. Andrzej Pigulski for many important comments and fruitful discussions that made this manuscript more comprehensible. This work has been supported by the Polish National Science Centre grant OPUS 16 No. 2018/31/B/ST9/00334. M.R. and I.S., P.K.S., P.I., M.G. acknowledge support from the Polish National Science Centre grant MAESTRO 8 No. 2016/22/A/ST9/00009, PRELUDIUM 18 No. 2019/35/N/ST9/03805, PRELUDIUM 18 No. 2019/35/N/ST9/02474, and the EU Horizon 2020 research and innovation program under grant agreement No 101004719, respectively. The authors made use of the Strasbourg astronomical data center (CDS). This research has made use of the SIMBAD database, operated at CDS, Strasbourg, France. This research has made use of the International Variable Star Index (VSX) database, operated at AAVSO, Cambridge, Massachusetts, USA.
	
	\software{TOPCAT \citep{2005ASPC..347...29T}, FNPEAKS (Z. Ko\l{}aczkowski, W. Hebisch, G. Kopacki), TATRY \citep{1996ApJ...460L.107S}}
	
	\clearpage
	\appendix
	\restartappendixnumbering
	\section{Data Tables}

	\begin{deluxetable*}{cccccllcc}[h]
	\centering
		\tabletypesize{\scriptsize}
		
		\tablecaption{Identification and Coordinates of HBSs}
		\label{tab:ident}

		\tablehead{\colhead{Loc} &\colhead{ID} & \colhead{Type} & \colhead{R.A.} & \colhead{Decl.} & \colhead{OGLE-IV ID} & \colhead{OGLE-III ID} & \colhead{OGLE-II ID} & \colhead{Other IDs} \\ 
		\colhead{} & \colhead{} & \colhead{} & \colhead{(h:m:s)} & \colhead{($\degree$:$'$:$''$)} & \colhead{} & \colhead{} & \colhead{} & \colhead{} }

		\startdata
		LMC & OGLE-LMC-HB-0001 & RG & 4:45:19.56 & -67:03:14.3 & LMC540.27.63 & LMC145.1.4667 & $-$ & $-$  \\
		LMC & OGLE-LMC-HB-0002 & RG & 4:46:24.84 & -72:47:57.8 & LMC528.21.4791 & $-$ & $-$ & OGLE-LMC-ECL-26648  \\
		LMC & OGLE-LMC-HB-0003 & MS & 4:47:02.42 & -68:55:01.7 & LMC539.08.103 & LMC141.8.8773 & $-$ & OGLE-LMC-ECL-26688  \\
		LMC & OGLE-LMC-HB-0004 & RG & 4:47:16.11 & -68:58:24.9 & LMC539.08.27 & LMC142.4.50 & $-$ & $-$  \\
		LMC & OGLE-LMC-HB-0005 & RG & 4:47:18.62 & -70:04:44.6 & LMC530.31.2375 & LMC143.1.18 & $-$ & $-$  \\
		LMC & OGLE-LMC-HB-0006 & RG & 4:47:24.27 & -70:40:39.1 & LMC530.14.7330 & LMC144.1.7335 & $-$ & $-$  \\
		LMC & OGLE-LMC-HB-0007 & RG & 4:47:27.76 & -69:29:08.2 & LMC531.15.2912 & LMC142.1.21 & $-$ & $-$  \\
		LMC & OGLE-LMC-HB-0008 & RG & 4:47:46.00 & -69:17:25.5 & LMC531.24.4 & LMC142.2.45 & $-$ & $-$  \\
		LMC & OGLE-LMC-HB-0009 & RG & 4:49:39.18 & -68:30:55.4 & LMC532.06.25753 & LMC141.3.13731 & $-$ & EROS2-star-lm015-5k-26494  \\
		LMC & OGLE-LMC-HB-0010 & RG & 4:50:08.58 & -70:01:04.8 & LMC530.29.22963 & LMC136.8.64 & $-$ & $-$ \\
		\vdots & \vdots &\vdots&\vdots&\vdots&\vdots&\vdots&\vdots & \vdots
		\enddata
		\tablecomments{Here we present the first 10 rows of the table. The full version of this table in a machine-readable format is provided. It is also available on the OGLE websites. Columns: (Loc) location indicator (GB—Galactic bulge, LMC—Large Magellanic Cloud, SMC—Small Magellanic Cloud); (ID) ID in the catalog which is OGLE-MMM-HB-NNNN, where MMM is a location indicator (BLG—Galactic bulge, LMC and SMC as above), and NNNN is a 4 digit number; (Type) Evolutionary status of the primary star, MS stands for stars evolving on the MS or post-MS and RG stands for stars evolving on the RGB/AGB or stars in the He-core-burning phase; (R.A.) Right ascension on J2000.0 epoch, in format (h:m:s); (Decl.) declination on J2000.0 epoch, in format ($\degree$:$'$:$''$); (OGLE-IV ID) OGLE-IV identifier of the star; (OGLE-III) OGLE-III identifier of the star; (OGLE-II) OGLE-II identifier of the star; (Others IDs) Catalog ID from OCVS or (if does not exist) the best match with SIMBAD/AAVSO/ASAS-SN database. Only a portion of this table is shown here to demonstrate its form and content. A machine-readable version of the full table is available.}
	\end{deluxetable*}
	
	\begin{deluxetable*}{cccccccccrlc}[h]
		\centering
		\tabletypesize{\scriptsize}
		
		\tablecaption{Basic Observational and Orbital Information about Each HBS System}
		\label{tab:hb}
		
		\tablehead{\colhead{Loc} &\colhead{ID} & \colhead{$I$} & \colhead{$V$} & \colhead{$P$} & \colhead{$T_0$} & \colhead{$A$} & \colhead{$e$} & \colhead{$i$} & \colhead{$\omega$} & \colhead{VAR} & \colhead{model} \\ 
			\colhead{} & \colhead{} & \colhead{(mag)} & \colhead{(mag)} & \colhead{(days)} & \colhead{(HJD$-$2,450,000)} & \colhead{(mag)} & \colhead{} & \colhead{(deg)} & \colhead{(deg)} & \colhead{} & \colhead{} } 
		
		\startdata
		LMC & OGLE-LMC-HB-0001 & 15.905 & 17.291 & 376.4172 & 9231.2363 & 0.030 & 0.3144 & 54.09 & 132.27 & $-$ & 1\\
		LMC & OGLE-LMC-HB-0002 & 16.520 & 17.900 & 207.3538 & 9015.7180 & 0.039 & 0.3138 & 79.73 & 24.28  & $-$ & 1\\
		LMC & OGLE-LMC-HB-0003 & 15.886 & 15.848 & 4.186218 & 9002.5750 & 0.030 & 0.1357 & 41.70 & 168.15 & $-$ & 1\\
		LMC & OGLE-LMC-HB-0004 & 15.574 & 17.283 & 243.9647 & 9112.6377 & 0.028 & 0.2913 & 30.71 & 173.36 & ECL & 0\\
		LMC & OGLE-LMC-HB-0005 & 14.225 & 15.753 & 494.5224 & 9437.2775 & 0.037 & 0.3109 & 61.98 & 120.44 & $-$ & 1\\
		LMC & OGLE-LMC-HB-0006 & 15.853 & 17.360 & 202.3257 & 9000.2190 & 0.029 & 0.1833 & 51.69 & 92.69  & TEO & 1\\
		LMC & OGLE-LMC-HB-0007 & 14.506 & 15.943 & 662.8316 & 9164.6504 & 0.021 & 0.5191 & 69.81 & 14.44  & $-$ & 1\\
		LMC & OGLE-LMC-HB-0008 & 14.271 & 15.623 & 404.1963 & 9264.7313 & 0.031 & 0.2117 & 50.54 & 2.64   & $-$ & 1\\
		LMC & OGLE-LMC-HB-0009 & 15.161 & 16.968 & 592.5599 & 9482.8130 & 0.061 & 0.4466 & 59.90 & 79.23  & $-$ & 1\\
		LMC & OGLE-LMC-HB-0010 & 15.094 & 16.287 & 177.1610 & 9062.7517 & 0.023 & 0.4071 & 36.65 & 65.79  & $-$ & 1\\
		\vdots&\vdots&\vdots&\vdots&\vdots&\vdots&\vdots&\vdots&\vdots&\vdots&\vdots&\vdots
		\enddata
		\tablecomments{Here we present the first 10 rows of the table. The full version of this table in a machine-readable format is provided. It is also available on the OGLE websites. Columns: (Loc) location indicator (GB—Galactic bulge, LMC—Large Magellanic Cloud, SMC—Small Magellanic Cloud); (ID) Catalog ID of HBS in the OCVS, description like for Table \ref{tab:ident}; ($I$) Cousins $I$-band mean magnitude; ($V$) Johnson $V$-band mean magnitude; ($P$) orbital period, in days; ($T_0$) the time of the periastron passage, in HJD$-$2,450,000; ($A$) the peak-to-peak amplitude of the $I$-band brightness variations, based on the fitted Kumar et al.'s model, in magnitude; ($e$) orbital eccentricity; ($i$) orbital inclination, in degrees; ($\omega$) argument of the periastron, in degrees; (VAR) additional variability; (model) flag indicating whether the HBS light curve exhibits proper Kumar et al.'s model (``1" is for proper model and ``0" is for improper model). Only a portion of this table is shown here to demonstrate its form and content. A machine-readable version of the full table is available.}
	\end{deluxetable*}	

	\begin{deluxetable*}{ccrrrrrrr}[h]
		\centering
		\tabletypesize{\scriptsize}
		
		\tablecaption{Kumar et al.'s Model Parameters from the MCMC Fitting Procedure (see Paper II, their Section 3)}
		\label{tab:kumar}
		
		\tablehead{\colhead{Loc} & \colhead{ID} & \colhead{$P$} & \colhead{$e$} & \colhead{$i$} & \colhead{$\omega$} & \colhead{$T_0$}& \colhead{$S$} & \colhead{$C$} \\ 
			\colhead{} & \colhead{} & \colhead{(days)} & \colhead{} & \colhead{(deg)} & \colhead{(deg)} & \colhead{(HJD$-$2,450,000)} & \colhead{} & \colhead{} } 
		
		\startdata
		LMC & OGLE-LMC-HB-0001 & $376.41_{\rm-2.6E-1}^{\rm+2.7E-1}$ & $0.314_{\rm-1.8E-2}^{\rm+1.8E-2}$ & $54.1_{\rm-2.9}^{\rm+3.6}$ & $132.3_{\rm-5.2}^{\rm+5.2}$ &  $9231.23_{\rm-4.38}^{\rm+4.19}$ & $0.005_{\rm-5.1E-4}^{\rm+5.2E-4}$ & $0.0013_{\rm-5.4E-4}^{\rm+5.5E-4}$ \\
		LMC & OGLE-LMC-HB-0002 & $207.353_{\rm-1.2E-1}^{\rm+1.2E-1}$ & $0.314_{\rm-2.0E-2}^{\rm+2.0E-2}$ & $79.7_{\rm-9.2}^{\rm+8.5}$ & $24.3_{\rm-5.5}^{\rm+5.5}$ &  $9015.72_{\rm-1.95}^{\rm+1.96}$ & $0.005_{\rm-4.8E-4}^{\rm+5.5E-4}$ & $0.0020_{\rm-7.4E-4}^{\rm+6.8E-4}$ \\
		LMC & OGLE-LMC-HB-0003 & $4.18622_{\rm-3.1E-5}^{\rm+2.8E-5}$ & $0.136_{\rm-1.3E-2}^{\rm+1.4E-2}$ & $41.7_{\rm-2.4}^{\rm+2.7}$ & $168.1_{\rm-6.4}^{\rm+6.3}$ & $9002.58_{\rm-0.07}^{\rm+0.07}$ & $0.015_{\rm-1.5E-3}^{\rm+1.5E-3}$ & $-0.0045_{\rm-1.5E-3}^{\rm+1.5E-3}$ \\
		LMC & OGLE-LMC-HB-0004 & $243.96_{\rm-9.4E-2}^{\rm+9.5E-2}$ & $0.291_{\rm-2.4E-2}^{\rm+2.7E-2}$ & $30.7_{\rm-1.4}^{\rm+1.4}$ & $173.4_{\rm-8.6}^{\rm+8.6}$ &  $9112.64_{\rm-2.68}^{\rm+2.46}$ & $0.010_{\rm-1.2E-3}^{\rm+1.2E-3}$ & $-0.0066_{\rm-8.7E-4}^{\rm+7.9E-4}$ \\
		LMC & OGLE-LMC-HB-0005 & $494.52_{\rm-2.8E-1}^{\rm+2.7E-1}$ & $0.311_{\rm-9.7E-3}^{\rm+9.7E-3}$ & $62.0_{\rm-2.8}^{\rm+3.4}$ & $120.4_{\rm-3.2}^{\rm+3.2}$ &  $9437.28_{\rm-2.74}^{\rm+2.74}$ & $0.006_{\rm-4.3E-4}^{\rm+4.4E-4}$ & $0.0005_{\rm-4.7E-4}^{\rm+4.6E-4}$ \\
		LMC & OGLE-LMC-HB-0006 & $202.32_{\rm-6.2E-2}^{\rm+6.3E-2}$ & $0.183_{\rm-1.1E-2}^{\rm+1.2E-2}$ & $51.7_{\rm-2.5}^{\rm+3.0}$ & $92.7_{\rm-4.0}^{\rm+4.1}$ &   $9000.22_{\rm-2.01}^{\rm+2.04}$ & $0.010_{\rm-8.6E-4}^{\rm+8.9E-4}$ & $-0.0017_{\rm-8.0E-4}^{\rm+7.9E-4}$ \\
		LMC & OGLE-LMC-HB-0007 & $662.83_{\rm-6.7E-1}^{\rm+6.8E-1}$ & $0.519_{\rm-2.2E-2}^{\rm+2.2E-2}$ & $69.8_{\rm-8.6}^{\rm+9.6}$ & $14.4_{\rm-6.4}^{\rm+6.7}$ &   $9164.65_{\rm-5.04}^{\rm+5.18}$ & $0.001_{\rm-1.7E-4}^{\rm+1.9E-4}$ & $0.0006_{\rm-3.4E-4}^{\rm+3.2E-4}$ \\
		LMC & OGLE-LMC-HB-0008 & $404.19_{\rm-2.4E-1}^{\rm+2.4E-1}$ & $0.212_{\rm-1.2E-2}^{\rm+1.2E-2}$ & $50.5_{\rm-2.0}^{\rm+2.3}$ & $2.6_{\rm-3.7}^{\rm+3.7}$ &    $9264.73_{\rm-3.67}^{\rm+3.72}$ & $0.010_{\rm-6.3E-4}^{\rm+6.3E-4}$ & $-0.0004_{\rm-6.8E-4}^{\rm+7.0E-4}$ \\
		LMC & OGLE-LMC-HB-0009 & $592.55_{\rm-1.7E-1}^{\rm+1.7E-1}$ & $0.447_{\rm-7.6E-3}^{\rm+7.6E-3}$ & $59.9_{\rm-1.7}^{\rm+2.0}$ & $79.2_{\rm-2.3}^{\rm+2.4}$ &   $9482.81_{\rm-1.67}^{\rm+1.66}$ & $0.006_{\rm-3.8E-4}^{\rm+3.8E-4}$ & $-0.0007_{\rm-4.0E-4}^{\rm+4.0E-4}$ \\
		LMC & OGLE-LMC-HB-0010 & $177.16_{\rm-4.5E-2}^{\rm+4.5E-2}$ & $0.407_{\rm-1.7E-2}^{\rm+1.6E-2}$ & $36.7_{\rm-0.9}^{\rm+0.9}$ & $65.8_{\rm-3.6}^{\rm+3.5}$ &   $9062.75_{\rm-1.11}^{\rm+1.12}$ & $0.006_{\rm-5.0E-4}^{\rm+5.3E-4}$ & $-0.0028_{\rm-4.5E-4}^{\rm+4.4E-4}$ \\
		\vdots&\vdots&\vdots&\vdots&\vdots&\vdots&\vdots&\vdots&\vdots
		\enddata

		\tablecomments{Here we present the first 10 rows of the table. The full version of this table in a machine-readable format is provided. Note that in the machine-readable format the uncertainties are written in separate columns, which are located next to the given parameter. Lower and upper indices denote the parameter uncertainties. They are 50th–16th and 84th-–50th percentile of parameter distribution from the MCMC modeling, respectively. Columns: (Loc) location indicator (GB—Galactic bulge, LMC—Large Magellanic Cloud, SMC—Small Magellanic Cloud); (ID) catalog ID of HBS in OCVS, description like for Table \ref{tab:ident}; ($P$, $\sigma_P^{-}$, $\sigma_P^{+}$) orbital period, its left- and right-side uncertainties, in days; ($e$, $\sigma_e^{-}$, $\sigma_e^{+}$) orbital eccentricity, its left- and right-side uncertainties; ($i$, $\sigma_i^{-}$, $\sigma_i^{+}$) orbital inclination, its left- and right-side uncertainties, in degrees; ($\omega_0$, $\sigma_{\omega}^{-}$, $\sigma_{\omega}^{+}$) argument of periastron, its left- and right-side uncertainties, in degrees; ($T_0$, $\sigma_{T_0}^{-}$, $\sigma_{T_0}^{+}$) epoch of the periastron passage, its left- and right-side uncertainties, in HJD$-$2,450,000; ($S$, $\sigma_S^{-}$, $\sigma_S^{+}$), amplitude scaling factor, its left- and right-side uncertainties; ($C$, $\sigma_C^{-}$, $\sigma_C^{+}$) zero-point offset, its left- and right-side uncertainties. Only a portion of this table is shown here to demonstrate its form and content. A machine-readable version of the full table is available.}
	\end{deluxetable*}
	\clearpage
	\bibliography{main}
	\bibliographystyle{aasjournal}
\end{document}